\documentclass[aps,prl,twocolumn,superscriptaddress,preprintnumbers,amsmath,amssymb,floatfix,nofootinbib]{revtex4}
\usepackage{graphicx}
\usepackage{dcolumn}
\usepackage{bm}
\usepackage{url}
\usepackage{amsmath}
\usepackage{amssymb}
\usepackage{color}
\usepackage{subfig}
\usepackage{ulem}

\begin{document}

\title{Challenging the K\"orner-Pati-Woo Theorem in Charmed Baryon Decays}

\author{Di Wang}
\affiliation{Department of Physics, Hunan Normal University, Changsha, 410081, China}

\author{Ying-Xin Lai}
\affiliation{Department of Physics, Hunan Normal University, Changsha, 410081, China}

\author{Fu-Sheng Yu}\email[Corresponding author: ]{yufsh@lzu.edu.cn}
\affiliation{Frontiers Science Center for Rare Isotopes, and School of Nuclear Science and Technology, Lanzhou University,
Lanzhou, 730000, China}
\affiliation{Lanzhou Center for Theoretical Physics, Key Laboratory of Theoretical Physics of Gansu Province, Key Laboratory of Quantum Theory and Applications of MoE, Gansu Provincial Research Center for Basic Disciplines of Quantum Physics, Lanzhou University, Lanzhou, 730000, China}

\begin{abstract}
The K\"orner-Pati-Woo (KPW) theorem is a long-standing selection rule widely used in heavy-baryon decays.
In this work, we argue that the conventional valence-color argument underlying the KPW theorem does not, by itself, establish an exact selection rule.
The $\rm KPW$-$\rm SU(3)_F$ prediction for $\mathcal{B}(\Xi_c^0\to pK^-)$, inferred from $\Lambda_c^+\to \Xi^0K^+$, is $17.2\pm6.0$ times the measured
value.
A joint likelihood fit to five observables in $\Xi_c^0\to pK^-$ and $\Lambda_c^+\to \Xi^0K^+$ yields a $6.3\sigma$ discrepancy under the $\rm KPW$-$\rm SU(3)_F$ ansatz.
A comparison of $\alpha(\Lambda_c^+\to \Delta^{++}K^-)$ and $\alpha(\Lambda_c^+\to \Sigma^{*}\pi)$ reveals a pronounced sign-sensitive tension.
Accommodating these measurements within the KPW constraint requires anomalously large $\rm SU(3)_F$ breaking effects.
Future measurements of $\Lambda^+_c\to\Sigma^{(*)0}K^+$, $\Sigma^{(*)+}K^0_S$, $\Delta^+\eta^{(\prime)}$, $\Delta^{++}\pi^-$, $\Delta^+\pi^0$, $\Delta^0\pi^+$, $\Delta^{++}K^-$, and $\Delta^{+}K^0_S$ would provide high-precision tests of the KPW theorem.

\end{abstract}

\maketitle

\section{Introduction}

Weak decays of heavy baryons probe heavy-to-light baryonic transitions and nonperturbative QCD dynamics.
The K\"orner-Pati-Woo (KPW) theorem \cite{Pati:1970fg,Korner:1970xq} states that two quarks produced at a weak vertex must be antisymmetric in flavor if both enter the same low-lying baryon.
This constraint is commonly imposed to reduce the topological or irreducible amplitudes.

In this work, we identify a limitation of the conventional argument for the KPW theorem: gluon exchange permits color configurations beyond the valence three-quark projection.
Predictions obtained by imposing the KPW constraint, notably for $\mathcal{B}(\Xi_c^0\to pK^-)$ and $\alpha(\Lambda_c^+\to \Delta^{++}K^-)$, are in substantial tension with measurements.
Reconciling these observables while preserving the KPW constraint requires very large $\rm SU(3)_F$ breaking effects.
We also identify branching fractions and decay asymmetries whose improved measurements can provide stringent tests of the KPW theorem.


\section{Theoretical argument}\label{kpw}

The KPW theorem \cite{Pati:1970fg,Korner:1970xq} requires two quarks produced by the weak operator to be antisymmetric in flavor when both enter the same low-lying baryon.
The conventional argument proceeds as follows.
Consider the four-quark operator
\begin{align}
\mathcal{O}(x)=&\{\overline{q}_{\alpha,i}(x)\gamma_\mu(1-\gamma_5)q_{\beta,j}(x)\}\nonumber\\&\qquad
\otimes\{\overline{q}_{\gamma,k}(x)\gamma^\mu(1-\gamma_5)q_{\delta,l}(x)\},
\end{align}
where $\alpha$, $\beta$, $\gamma$, and $\delta$ denote color indices, $i$, $j$, $k$, and $l$ denote flavor indices.
A Fierz rearrangement gives
\begin{align}
\mathcal{O}(x)=&\{\overline{q}_{\gamma,k}(x)\gamma_\mu(1-\gamma_5)q_{\beta,j}(x)\}\nonumber\\&\qquad
\otimes\{\overline{q}_{\alpha,i}(x)\gamma^\mu(1-\gamma_5)q_{\delta,l}(x)\}.
\end{align}
Thus, $\mathcal{O}(x)$ is invariant under $(\alpha,i) \leftrightarrow (\gamma,k)$.
If both quarks enter the final-state baryon, the valence color-singlet projection selects their color-antisymmetric component.
Together with the interchange symmetry, this leads to an antisymmetric flavor combination.

\begin{figure}[t!]
  \centering
  \includegraphics[width=7.5cm]{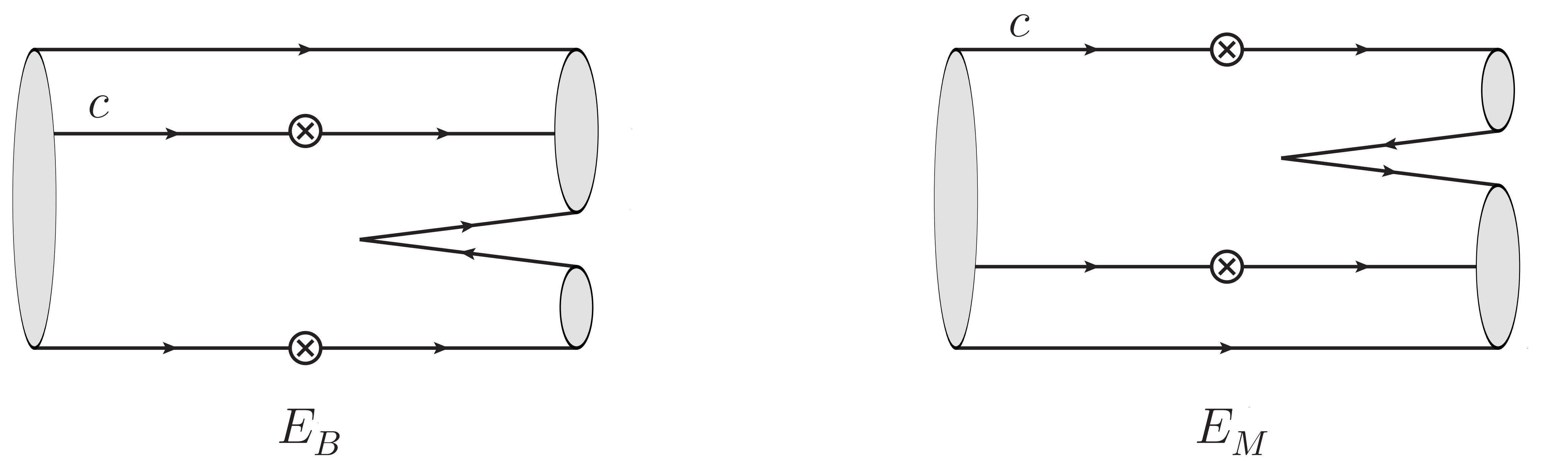}
  \caption{Representative diagrams contributing to $\Lambda^+_c\to \Xi^0K^+$ (left) and $\Xi^0_c\to pK^-$ (right).}\label{f3}
\end{figure}
\begin{figure}[t!]
  \centering
  \includegraphics[width=8.6cm]{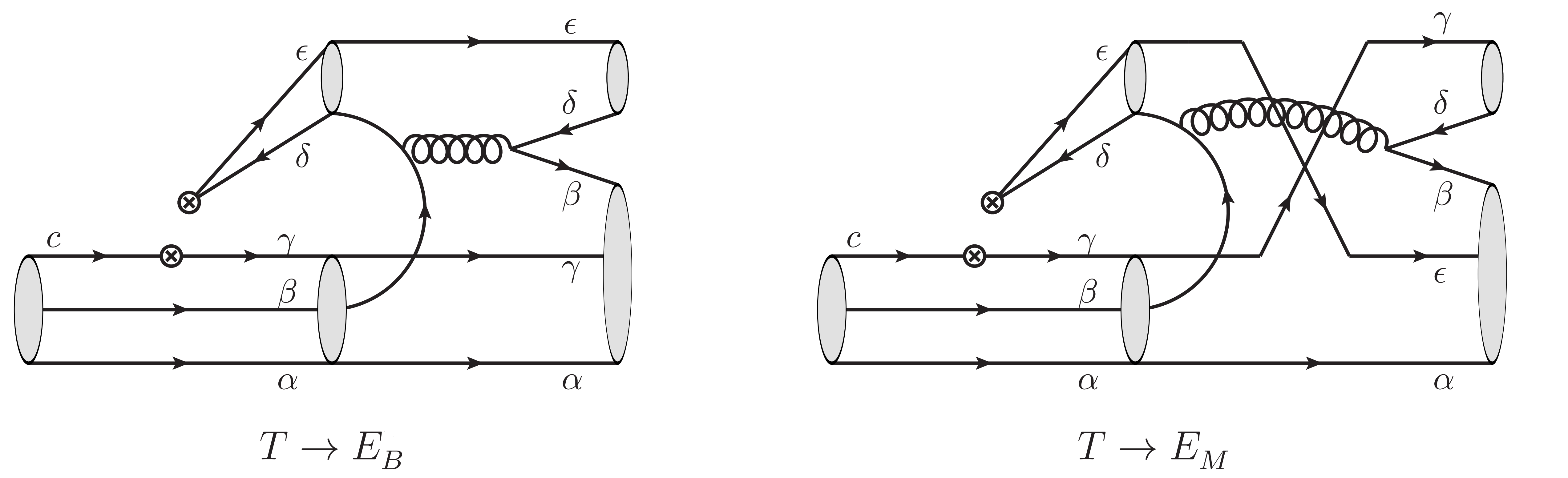}
  \caption{Schematic rescattering contributions in diagrams $E_B$ and $E_M$.}\label{f4}
\end{figure}

Combining the KPW constraint with flavor symmetry yields amplitude relations beyond those implied by flavor symmetry alone.
For instance, neglecting terms proportional to $\lambda_b$, the KPW constraint together with $\rm SU(3)_F$ symmetry implies \cite{Zhong:2024qqs,Geng:2019xbo,Wang:2024ztg}
\begin{align}\label{eq:kpw}
\frac{\mathcal{A}(\Lambda_c^+\to \Xi^0K^+)}{V_{cs}^*V_{ud}}=-\frac{\mathcal{A}(\Xi_c^0\to pK^-)}{V_{cs}^*V_{us}}.
\end{align}
However, this relation is in tension with the rescattering description \cite{Lai:2025ryi}, as illustrated by the topological analysis below.
Fig.~\ref{f3} shows representative diagrams for $\Lambda_c^+\to \Xi^0K^+$ and $\Xi_c^0\to pK^-$, respectively.
Under the KPW constraint, $E_B$ and $E_M$ satisfy $E_B=-E_M$ \cite{Wang:2024ztg,Zhong:2024qqs}.
The long-distance rescattering contributions to $E_B$ and $E_M$ are shown in Fig.~\ref{f4}.
For the color flows depicted in Fig.~\ref{f4}, the final-state color-singlets imply $\delta=\epsilon$ in $T\to E_B$ and $\gamma=\delta=\epsilon$ in $T\to E_M$.
Consequently, $E_B$ is color favored whereas $E_M$ is color suppressed.

We now examine the color projection underlying the conventional KPW argument.
The argument projects the two quarks created at the weak vertex onto the color-antisymmetric pair in a valence three-quark baryon \cite{Pati:1970fg,Korner:1970xq}.
This projection, however, does not exhaust the color configurations allowed by QCD.
Gluon exchange can redistribute color among the participating quarks.
Although a pure three-quark color singlet requires any two quarks to form a color-antitriplet, this restriction does not apply to a three-quark subsystem embedded in a larger color singlet.
For example,  a symmetric color-sextet quark pair can combine with a third quark to form a color-octet subsystem, which may couple to an additional gluon to produce an overall singlet.
Such configurations naturally arise in the multiquark color space of baryon-meson final states and in higher Fock components containing gluons.
The color-singlet nature of an asymptotic baryon does not, by itself, require the two quarks produced at the weak vertex to remain in a color-antitriplet configuration throughout hadronization.
Thus, the KPW theorem should be treated as a conditional constraint rather than an exact conclusion.
Its applicability remains to be tested experimentally.

\section{Experimental test}\label{mo}

The partial-wave amplitudes of $\Lambda_c^+\to \Xi^0K^+$ are constrained by BESIII measurements, which, together with the LHCb measurements for $\Xi_c^0\to pK^-$, enables a direct test of the KPW theorem.
We define the ratio of the branching fraction for $\Xi_c^0\to pK^-$ inferred from $\Lambda_c^+\to \Xi^0K^+$ to its measured value as
\begin{align}\label{gap}
  R_{\Xi_c^0\to pK^-}^{\rm KPW}\equiv\frac{\mathcal{B}(\Xi_c^0\to pK^-)|_{\Lambda_c^+\to \Xi^0K^+,\rm KPW}}{\mathcal{B}(\Xi_c^0\to pK^-)|_{\rm EXP}}.
\end{align}
According to Eq.~\eqref{eq:kpw}, $R_{\Xi_c^0\to pK^-}^{\rm KPW}$ can be calculated by
\begin{align}
  &R_{\Xi_c^0\to pK^-}^{\rm KPW}=\frac{\mathcal{B}(\Lambda_c^+\to \Xi^0K^+)}{\mathcal{B}(\Xi_c^0\to pK^-)}\frac{\tau_{\Xi_c^0}}{\tau_{\Lambda_c^+}}
\frac{\Phi_{pK^-}}{\Phi_{\Xi^0K^+}}\left|\frac{V_{us}}{V_{ud}}\right|^2\nonumber\\&
\qquad\qquad\times \left[1+\left(\frac{\kappa^2_{pK^-}}{\kappa^2_{\Xi^0K^+}}-1\right)
 f_P^{\Xi^0K^+}\right],
\end{align}
where
\begin{align}
\Phi=p_c\frac{(m_{B_{c}}+m_B)^2-m_M^2}{m_{B_{c}}^2},\quad
\kappa=\frac{p_c}{E_B+m_B}.
\end{align}
$f_P^{\Xi^0K^+}$ is the $P$-wave fraction in the total decay rate,
\begin{align}\label{eq:fpdef}
 f_P^{\Xi^0K^+}
 =\frac{\kappa^2_{\Xi^0K^+}|\mathcal P_{\Xi^0K^+}|^2}
 {|\mathcal S_{\Xi^0K^+}|^2+
 \kappa^2_{\Xi^0K^+}|\mathcal P_{\Xi^0K^+}|^2}.
\end{align}
It can be expressed by the Lee-Yang parameter $\alpha_{\Xi^0K^+}$ and the helicity-amplitude phase $\Delta_{\Xi^0K^+}$,
\begin{align}\label{eq:fpalphadelta}
 f_P^{\Xi^0K^+}&=\frac{1-\gamma_{\Xi^0K^+}}{2}
 \nonumber\\&=\frac{1-\sqrt{1-\alpha_{\Xi^0K^+}^2}\cos\Delta_{\Xi^0K^+}}{2}.
\end{align}
BESIII measurements favor the $P$-wave-dominated configuration in $\Lambda_c^+\to \Xi^0K^+$ since $\gamma_{\Xi^0K^+}<0$, although an $S$-wave-dominated solution is also reported \cite{BESIII:2023wrw}.
With the following inputs \cite{BESIII:2023wrw,LHCb:2026wcy},
\begin{align}\label{eq:helgamma}
\mathcal{B}(\Lambda_c^+\to \Xi^0K^+)&=(5.5\pm0.7)\times10^{-3},\nonumber\\
\mathcal{B}(\Xi_c^0\to pK^-)&=(4.50\pm1.05)\times10^{-5},\nonumber\\
\alpha(\Lambda_c^+\to \Xi^0K^+)&=0.01\pm0.16\pm0.03,\nonumber\\
\Delta(\Lambda_c^+\to \Xi^0K^+)&=3.84\pm0.90\pm0.17\,\,\,\mathrm{rad},
\end{align}
$R_{\Xi_c^0\to pK^-}^{\rm KPW}$ is given by
\begin{align}\label{gap1}
  R_{\Xi_c^0\to pK^-}^{\rm KPW}=17.2\pm6.0.
\end{align}
This discrepancy is difficult to accommodate with conventional estimates of $\rm SU(3)_F$-breaking effects at the $20\%-30\%$ level.
To reduce sensitivity to uncertainties in the angular observables, we also derive the range allowed by the branching fractions alone,
\begin{align}
  R_{\Xi_c^0\to pK^-}^{\rm KPW}\in \left[5.6\pm1.5,\,18.7\pm 5.0\right].
\end{align}
This range indicates substantial tension between the observed
branching fractions and the KPW-$\rm SU(3)_F$ hypothesis, even without angular information.

\begin{table}[t!]
\centering
\caption{Experimental inputs, fitted values, and standardized residuals in the joint likelihood fit.}
\label{JL}
\small
\renewcommand{\arraystretch}{1.08}
\begin{tabular}{lccc}
\hline\hline
Observable \,& \,Central value \,& \,Best fit \,& \,Pull\\\hline

$\mathcal{B}(\Lambda_c^+\to \Xi^0K^+)\;(10^{-3})$ & 5.50 & 1.38 & $-5.89$\\
$\alpha(\Lambda_c^+\to \Xi^0K^+)$ & $0.01$ & $0.13$ & $+0.76$\\
$\Delta(\Lambda_c^+\to \Xi^0K^+)\;(\mathrm{rad})$ & 3.84 & 6.09 & $+2.46$\\
$\mathcal{B}(\Xi_c^0\to pK^-)\;(10^{-5})$ & 4.50 & 6.47 & $+1.88$\\
$\alpha(\Xi_c^0\to pK^-)$ & $0.32$ & $0.24$ & $-0.55$\\\hline
\hline
\end{tabular}
\end{table}
We perform a joint likelihood fit to the observables in Eq.~\eqref{eq:helgamma}, supplemented by $\alpha(\Xi_c^0\to pK^-)=0.32\pm0.15\pm0.01$ \cite{LHCb:2026wcy}.
The fit yields $\chi^2_{\min}=45.2$ for two nominal degrees of freedom, reveals a $6.3\sigma$ discrepancy under the KPW-$\rm SU(3)_F$ ansatz.
Table~\ref{JL} lists the fitted values and pulls.
To reduce the predicted \(\Xi_c^0\to pK^-\) rate, the fit is forced to suppress $\mathcal{B}(\Lambda^+_c\to \Xi^0K^+)$ from its experimental central value of \(5.50\times10^{-3}\) down to \(1.38\times10^{-3}\), corresponding to a $5.89\sigma$ pull.

\begin{figure*}[t!]
\begin{minipage}[c]{0.95\textwidth}
  \centering
  \includegraphics[width=8cm]{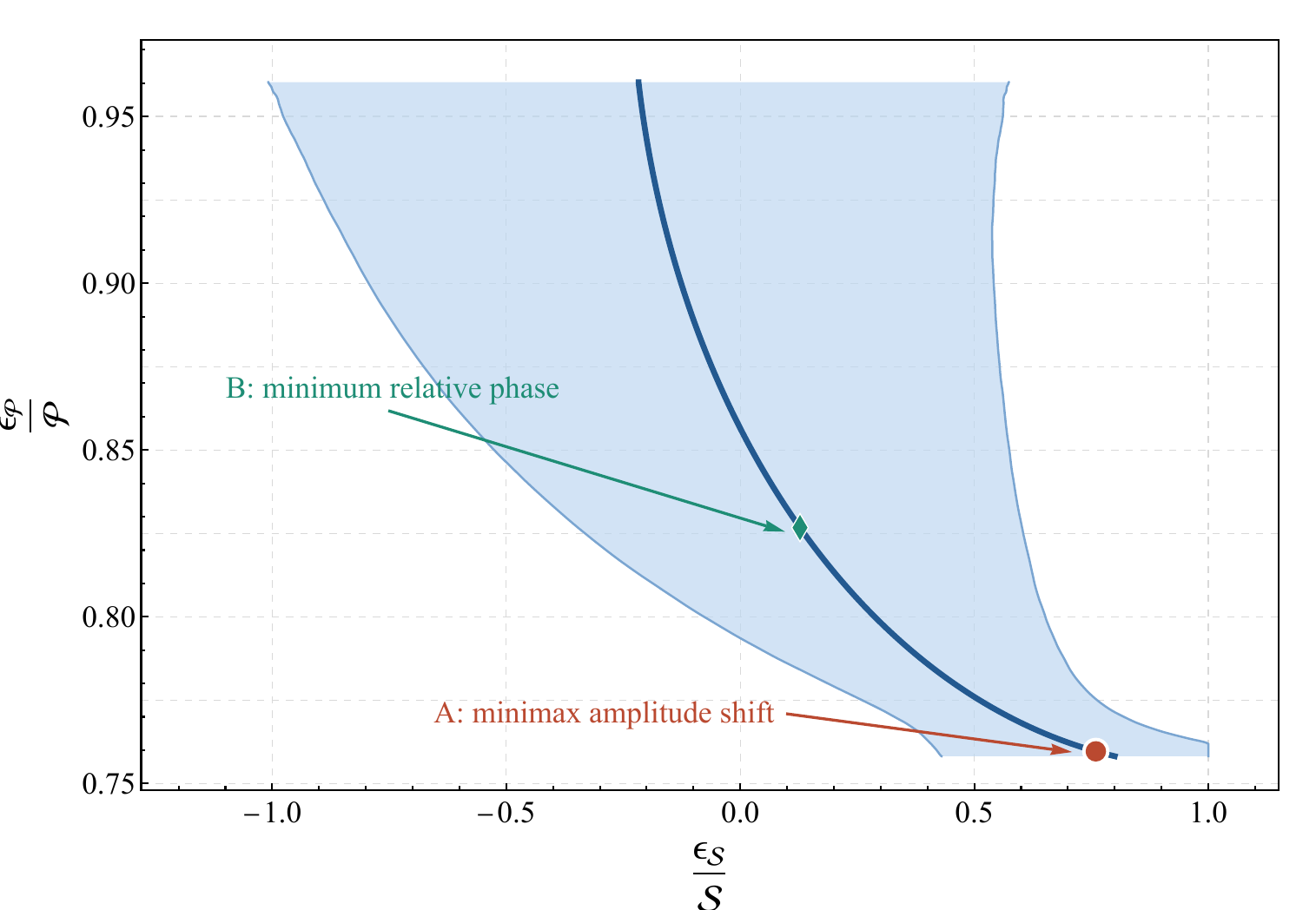}\qquad
  \includegraphics[width=8cm]{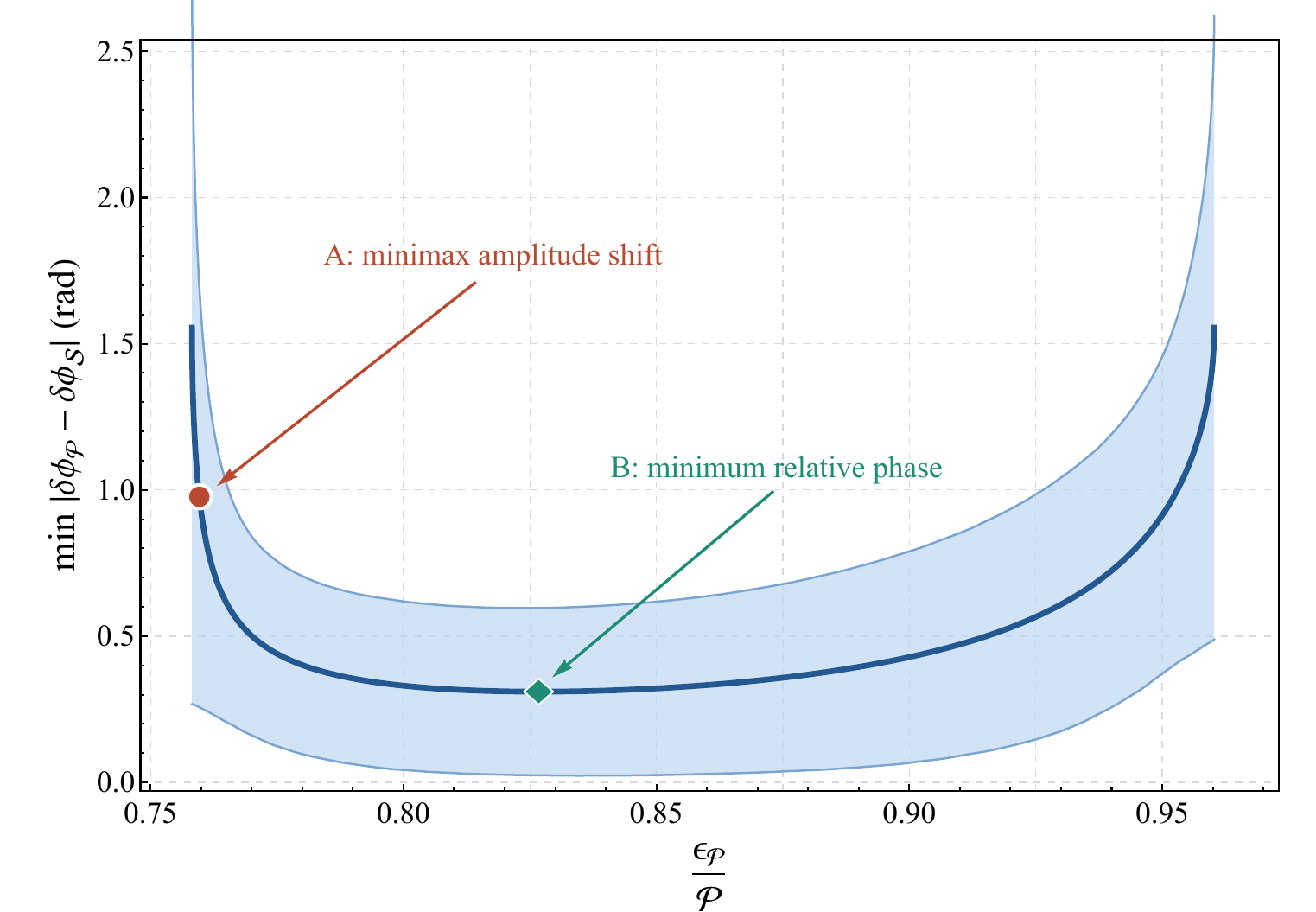}
\end{minipage}
  \caption{Relative $S$- and $P$-wave magnitude shifts (left) and minimal relative phase shift versus of $P$-wave magnitude shift (right).
  Points A and B minimize $\max\!\left(\left|\epsilon_S/\mathcal{S}\right|,\left|\epsilon_P/\mathcal{P}\right|\right)$ and $\min\!|\delta\phi_P-\delta\phi_S|$, respectively.
  Shaded regions denote uncertainty bands. }\label{f1}
\end{figure*}

To quantify $\rm SU(3)_F$-breaking required to maintain the KPW constraint, we parameterize the $S$- and $P$-wave amplitudes in $\Lambda_c^+\to \Xi^0K^+$ and $\Xi_c^0\to pK^-$ as
\begin{align}
  \mathcal{S}_{\Xi^0K^+}&=\mathcal{S},\quad \mathcal{S}_{pK^-}=\lambda(\mathcal{S-\epsilon_S})e^{i\delta\phi_S}, \nonumber\\ \mathcal{P}_{\Xi^0K^+}&=\mathcal{P}e^{i\phi_P},
 \quad \mathcal{P}_{pK^-}=\lambda(\mathcal{P-\epsilon_P})e^{i(\phi_P+\delta\phi_P)},
\end{align}
with $\lambda=|V_{us}/V_{ud}|$.
In this model-independent ansatz, only the relative phase $\delta\phi_P-\delta\phi_S$ is observable since a common shift of $\delta\phi_S$ and $\delta\phi_P$ is a pure rephasing.
We characterize the solution manifold determined by the five observables.
The relative $S$- and $P$-wave magnitude shifts and the minimum relative phase shift versus the $P$-wave shift are shown in Fig.~\ref{f1}\footnote{
The uncertainty bands also contain regions with near-unity values of \(\epsilon_P/\mathcal{P}\) (\(\epsilon_S/\mathcal{S}\)) and infinite \(\epsilon_S/\mathcal{S}\) (\(\epsilon_P/\mathcal{P}\)).
The reason is that, for the $P$($S$)-wave dominated solution of $\Lambda_c^+\to \Xi^0K^+$, the uncertainty interval of the $S$($P$)-wave amplitude includes zero \cite{BESIII:2023wrw}, hence \(\epsilon_S/\mathcal{S}\) (\(\epsilon_P/\mathcal{P}\)) becomes singular.}.
The solutions require substantial amplitude-level modifications.
The minimax point of the relative $S$- and $P$-wave magnitude shifts is $\max\!\left(\left|\epsilon_S/\mathcal{S}\right|,\left|\epsilon_P/\mathcal{P}\right|\right)_{\rm min} =76.0\%$, with relative phase shift ${\rm min}|\delta\phi_P-\delta\phi_S|=56.0^{\circ}$.
The minimum-phase point is ${\rm min}|\delta\phi_P-\delta\phi_S|=17.8^\circ$, with $\epsilon_P/\mathcal{P} = 82.7\%$.

The four $\Xi^0_c$ decay modes $\Sigma^+K^-$, $pK^-$, $\Sigma^+\pi^-$, and $p\pi^-$ form a $U$-spin system.
In the U-spin limit, neglecting the small $\lambda_b$ terms, their amplitudes are related by CKM factors.
Analogously to Eq.~\eqref{gap}, we give the ratio of the KPW prediction for $\mathcal{B}(\Xi_c^0\to \Sigma^+K^-)$, inferred from $\Lambda_c^+\to \Xi^0K^+$, to its measured value,
\begin{align}\label{gap2}
  R_{\Xi_c^0\to \Sigma^+K^-}^{\rm KPW}&\equiv\frac{\mathcal{B}(\Xi_c^0\to \Sigma^+K^-)|_{\Lambda_c^+\to \Xi^0K^+,\rm KPW}}{\mathcal{B}(\Xi_c^0\to \Sigma^+K^-)|_{\rm EXP}}\nonumber\\
  &=4.96\pm 1.44.
\end{align}
By contrast, the U-spin prediction obtained from $\Xi_c^0\to pK^-$ gives the ratio as
\begin{align}\label{u}
  R_{\Xi_c^0\to \Sigma^+K^-}^{\rm U}&\equiv\frac{\mathcal{B}(\Xi_c^0\to \Sigma^+K^-)|_{\Xi^0_c\to pK^-,\rm U}}{\mathcal{B}(\Xi_c^0\to \Sigma^+K^-)|_{\rm EXP}}\nonumber\\
  &\in [0.29\pm0.09,\,0.48\pm 0.15].
\end{align}
This ratio is smaller than the KPW-based prediction.
The U-spin breaking implied by the theory-experiment discrepancy in Eq.~\eqref{u} is plausible.
Differences between $s$- and $d$-quark loop contributions might enhance U-spin breaking in $\Xi^0_c\to \{\Sigma^+K^-, pK^-, \Sigma^+\pi^-, p\pi^-\}$, analogous to mechanism in $D^0\to \{\pi^+K^-, K^+K^-, \pi^+\pi^-, K^+\pi^-\}$ \cite{Brod:2012ud,Hiller:2012xm}.
The same enhancement mechanism does not apply to the two Cabibbo-favored channels in Eq.~\eqref{gap2}.
Experimentally, $\mathcal{B}(\Xi_c^0\to \Sigma^+K^-)$ is inferred from a ratio normalized to $\mathcal{B}(\Xi^0_c\to \Xi^-\pi^+)$ \cite{Belle:2021avh}.
We favor the branching fraction of the reference channel given by PDG, \(\mathcal{B}(\Xi^0_c\to \Xi^-\pi^+)=(1.43 \pm 0.27) \times 10^{-2}\) \cite{PDG}, despite many theoretical studies predict a larger value.

Ref.~\cite{Lai:2025ryi} presents global fits to charmed-baryon decays into an octet baryon and a pseudoscalar meson in the $\rm SU(3)_F$ limit, both with and without the KPW constraint.
With the KPW constraint imposed, the best fit yields $\chi^2/{\rm d.o.f}\geq10.8$.
Without this constraint, the best fit yields $\chi^2/{\rm d.o.f}=2.6$.
Relaxing the KPW constraint also improves the description of several branching fractions and asymmetry parameters, notably those for $\Xi_c^0$ decay modes. 


\begin{table}[t!]
 \centering
 \caption{Deviations between the measured $\alpha(\Lambda_c^+\to\Delta^{++} K^-)$ and the KPW-$\rm SU(3)_F$  prediction.}
 \label{tab:pulls}
 \begin{tabular}{@{}llll@{}}
 \hline\hline
  $\alpha_{\rm EXP}$& Theory edge~~~ & $\Delta\alpha\pm\sigma$\qquad~~~ & $|\Delta\alpha|/\sigma$\\\hline
 $0.55\pm 0.04$~~~ & Lower & $-1.45\pm0.07$ & $19.9$\\
                  & Upper & $-1.37\pm0.08$ & $16.2$\\\hline\hline
 \end{tabular}
\end{table}
\begin{figure}[t!]
  \centering
  \includegraphics[width=7cm]{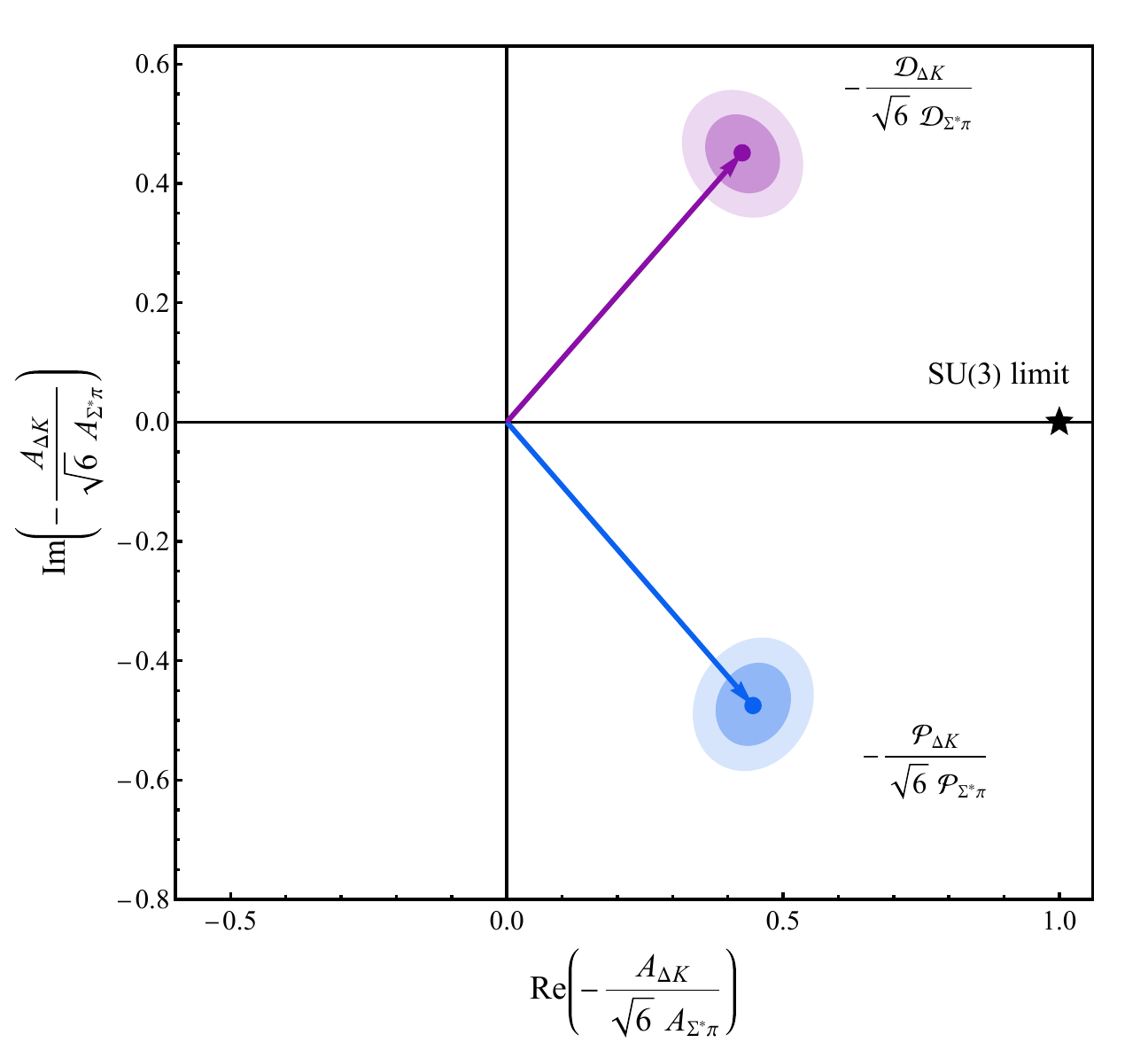}
  \caption{Minimum $\rm SU(3)_F$-breaking solution, where darker and lighter ellipses denote $68\%$ and $95\%$ confidence regions, respectively. }\label{fx}
\end{figure}
Charmed baryon decays into decuplet baryons provide complementary tests of the KPW theorem.
Isospin symmetry relates $\Lambda_c^+\to\Sigma^{*+}\pi^0$ and $\Lambda_c^+\to\Sigma^{*0}\pi^+$.
The branching fractions and $\alpha$ parameters of these two modes are given by \cite{BESIII:2022udq}:
\begin{align}\label{eq:BESdata}
 \mathcal{B}(\Lambda_c^+\to\Sigma^{*+}\pi^0)&=(5.86\pm0.49\pm0.52\pm0.36)\times10^{-3},\nonumber\\
  \mathcal{B}(\Lambda_c^+\to\Sigma^{*0}\pi^+)&=(6.47\pm0.59\pm0.66\pm0.38)\times10^{-3},\nonumber\\
 \alpha(\Lambda_c^+\to\Sigma^{*+}\pi^0)&=-0.917\pm0.069\pm0.056,\nonumber\\
 \alpha(\Lambda_c^+\to\Sigma^{*0}\pi^+)&=-0.789\pm0.098\pm0.056.
\end{align}
Averaging the two isospin-related measurements gives
\begin{align}
\overline{\mathcal{B}}(\Lambda_c^+\to\Sigma^{*}\pi)&=(6.10\pm0.66)\times10^{-3},\nonumber\\
 \overline\alpha(\Lambda_c^+\to\Sigma^{*}\pi)&=-0.87\pm0.07.
\end{align}
The KPW constraint, combined with $\rm SU(3)_F$ symmetry, implies \cite{Geng:2019awr,Wang:2024nxb}
\begin{align}\label{kpw2}
 \mathcal{A}(\Lambda_c^+\to\Delta^{++} K^-)
 =-\sqrt{6}\,\mathcal{A}(\Lambda_c^+\to\Sigma^{*}\pi).
\end{align}
Then $\mathcal{B}(\Lambda_c^+\to\Delta^{++} K^-)$ and $\alpha(\Lambda_c^+\to\Delta^{++} K^-)$ can be predicted from $\Lambda_c^+\to\Sigma^{*}\pi$,
\begin{align}
\mathcal{B}^{\rm KPW}_{\Lambda_c^+\to\Delta^{++} K^-}&\in[(4.27\pm0.47)\%,\,(4.68\pm0.51)\%],\nonumber\\
 \alpha^{\rm KPW}_{\Lambda_c^+\to\Delta^{++} K^-}&\in[-0.91\pm0.06,\,-0.83\pm0.07].
\end{align}
The PDG values \cite{PDG}, based on LHCb analyses \cite{LHCb:2022sck,LHCb:2023crj}, are given by
\begin{align}
\mathcal{B}(\Lambda_c^+\to\Delta^{++} K^-)
 &=(1.80\pm0.08)\%,\nonumber\\
 \alpha(\Lambda_c^+\to\Delta^{++} K^-)
 &=0.55\pm0.04.
 \label{eq:PDG}
\end{align}
Table~\ref{tab:pulls} compares the measured $\alpha$ with the KPW prediction.
The measurement and prediction differ markedly.
If the discrepancy is attributed to $\rm SU(3)_F$ breaking, we minimize the largest relative complex-amplitude correction, defined as
\begin{align}
 \eta\equiv\max\!\left(\left|-\frac{\mathcal{P}_{\Delta K}}{\sqrt{6}\,\mathcal{P}_{\Sigma^*\pi}}-1\right|,
\left|-\frac{\mathcal{D}_{\Delta K}}{\sqrt{6}\,\mathcal{D}_{\Sigma^*\pi}}-1\right|\right).
\end{align}
The solution is shown in Fig.~\ref{fx}.
The sign reversal requires a relative $P-D$ strong-phase shift of
at least \(93.5^\circ\).


Within the isospin symmetry, the KPW theorem leads to the following amplitude relations \cite{Wang:2025bdl,Geng:2019xbo,Wang:2024nxb}:
\begin{align}\label{b8}
 \mathcal{A}(\Lambda_c^+\to \Sigma^{(*)+}K^0) &= \sqrt{2}\,\mathcal{A}(\Lambda_c^+\to \Sigma^{(*)0}K^+),\nonumber\\
\mathcal{A}(\Lambda_c^+\to\Delta^+\eta) &= \mathcal{A}(\Lambda_c^+\to\Delta^+\eta^\prime)=0,\nonumber\\
\mathcal{A}(\Lambda^+_c\to \Delta^{++}\pi^-) &= -\sqrt{3}\mathcal{A}(\Lambda^+_c\to \Delta^{0}\pi^+)\nonumber\\&=-\sqrt{\frac{3}{2}}\mathcal{A}(\Lambda^+_c\to \Delta^{+}\pi^0),\nonumber\\
 \mathcal{A}(\Lambda_c^+\to \Delta^{+}K^0) &= \mathcal{A}(\Lambda_c^+\to \Delta^{0}K^+)=0,\nonumber\\
 \mathcal{A}(\Lambda_b^0\to\Sigma^{(*)-}K^+)&=-\sqrt{2}\mathcal{A}(\Lambda_b^0\to\Sigma^{(*)0}K^0),\nonumber\\
\mathcal{A}(\Lambda_b^0\to \Delta^{0}\eta)&=\mathcal{A}(\Lambda_b^0\to \Delta^{0}\eta^\prime)=0,\nonumber\\
\mathcal{A}(\Lambda_b^0\to \Delta^{-}\pi^+)&=-\sqrt{3}\mathcal{A}(\Lambda_b^0\to \Delta^{+}\pi^-)\nonumber\\&=\sqrt{\frac{3}{2}}\mathcal{A}(\Lambda_b^0\to \Delta^{0}\pi^0).
\end{align}
These relations provide high-precision tests of the KPW theorem, as isospin breaking is generally predicted to be $\mathcal{O}(1\%)$.
Octet-baryon modes can be reconstructed via electromagnetic or weak decays of the daughter particles, and decuplet-baryon modes can be isolated by amplitude analyses of the corresponding three-body final states.
In particular, a nonzero resonant amplitude for $\Lambda^+_c\to \Delta^+\eta^{(\prime)}$ or $\Lambda^0_b\to \Delta^0\eta^{(\prime)}$ would challenge the KPW theorem.
The doubly Cabibbo-suppressed mode $\Lambda_c^+\to \Delta^{+}K^0$ can be identified by measuring the difference $6\mathcal{B}(\Lambda_c^+\to \Delta^{+}K^0_S)-\mathcal{B}(\Lambda_c^+\to \Delta^{++}K^-)$, since $\mathcal{A}(\Lambda_c^+\to \Delta^{++}K^-)=\sqrt{3}\mathcal{A}(\Lambda_c^+\to \Delta^{+}\overline K^0)$ in the isospin limit and $\Lambda_c^+\to \Delta^{+}K^0$ contributes to $\Lambda_c^+\to \Delta^{+}K^0_S$.
With $\mathcal{B}(\Lambda_c^+\to \Delta^{++}K^-)=(1.80\pm0.08)\%$ \cite{PDG}, $\mathcal{B}(\Lambda_c^+\to \Delta^{+}K^0_S)$ is predicted to be $(2.99\pm0.13)\times 10^{-3}$ under the KPW-isospin ansatz.
The resonant $\Delta^{+}K^0_S$ contribution can be extracted from amplitude analyses of $\Lambda^+_c\to pK^0_S\pi^0$ and $\Lambda^+_c\to nK^0_S\pi^+$, and the later gives that $\mathcal{B}(\Lambda_c^+\to\Delta^{+}K^0_S)< 2.54\times 10^{-3}$ \cite{BESIII:2026zyf}.
Besides, the relation $\mathcal{A}(\Lambda^+_c\to \Delta^{++}\pi^-) = -\sqrt{3}\mathcal{A}(\Lambda^+_c\to \Delta^{0}\pi^+)$ can be tested in an amplitude analysis of $\Lambda^+_c\to p\pi^+\pi^-$.


\section{Summary}

In this work, we identify the conventional valence-color argument of the KPW theorem would be violated by gluon exchanges.
Predictions obtained under the KPW-$\rm SU(3)_F$ ansatz are in substantial tension with recent measurements.
Reproducing these measurements within the KPW constraint requires anomalously large $\rm SU(3)_F$ breaking effects.
Future measurements of $\Lambda^+_c\to \Sigma^{(*)0}K^+$, $\Sigma^{(*)+}K^0_S$, $\Delta^+\eta^{(\prime)}$, $\Delta^{++}\pi^-$, $\Delta^+\pi^0$, $\Delta^0\pi^+$, $\Delta^{++}K^-$, and $\Delta^{+}K^0_S$ would provide high-precision tests of the KPW theorem.

\begin{acknowledgements}

We are grateful to Wei Shan, Long-Ke Li, and Pei-Rong Li for useful discussions.
This work was supported in part by the National Natural Science Foundation of China under Grant No.~12335003 and No.~12247101, the Fundamental Research Funds for the Central Universities under Grant No.~lzujbky-2023-stlt01, lzujbky-2024-oy02, lzujbky-2025-jdzx07, and lzujbky-2025-eyt01, the Scientific Research Innovation Capability Support Project for Young Faculty under Grant No.~ZYGXQNJSKYCXNLZCXMP2, the Natural Science Foundation of Gansu Province under Grant No.~25JRRA799, and the "111 Center" under Grant No.~B20063, and the Scientific Research Fund of Hunan Provincial Department under Grant No.~25B0090.

\end{acknowledgements}

\end{document}